\documentclass[a4paper,11pt]{article}
\usepackage{pos}
\usepackage[symbol]{footmisc}
\renewcommand{\thefootnote}{\fnsymbol{footnote}}
\newcommand{\lsim} {\buildrel < \over {_\sim}}
\newcommand{\gsim} {\buildrel > \over {_\sim}}

\title{Probing secret interactions of eV-scale sterile neutrinos with the diffuse supernova neutrino background}
 \ShortTitle{Probing secret interactions with the diffuse supernova neutrino background}

\author*[a]{Mary Hall Reno}
\author[a,b]{Yu Seon Jeong}
\author[c]{Sergio Palomares-Ruiz}
\author[d]{Ina Sarcevic}

\affiliation[a]{Department of Physics and Astronomy, University of Iowa\\
  Iowa City, Iowa 52242, USA}

\affiliation[b]{Theory Division, CERN\\
1211 Geneve 23, Switzerland}

\affiliation[c]{Instituto de F\'isica Corpuscular (FIC), CSIC-Universitat de Val\`encia\\
E-46071 Val\`encia, Spain}

\affiliation[d]{Department of Physics, University of Arizona\\
Tucson, AZ 85721, USA}

\emailAdd{mary-hall-reno@uiowa.edu}
\emailAdd{yuseon.jeong@cern.ch}
\emailAdd{sergiopr@ific.uv.ed}
\emailAdd{ina@physics.arizona.edu}

\abstract{
While three flavors of ``active'' neutrinos are consistent with mixing angle results within error bars, there are anomalies may be hints of physics beyond the standard model that can accommodate a fourth mostly ``sterile" neutrino species with an eV-scale mass and a mixing angle with active neutrinos of order $\theta_0\simeq 0.1$. We describe a scenario with
eV-scale sterile neutrinos that have self-interactions via a new gauge vector boson, a ``secret'' mediator $\phi$. We show that their production in the early Universe via mixing with active neutrinos is consistent with Big Bang Nucleosynthesis and free-streaming constraints in the Cosmic Microwave Background epoch. For $M_\phi=4-8$ keV and sterile neutrino coupling $g_s=10^{-4}$,  we find that resonant interactions of diffuse supernova neutrinos with relic sterile neutrinos in transit to the Earth would cause spectral dips in the neutrino flux. We illustrate the corresponding (anti-)neutrino event distributions as a function of energy in the DUNE (Hyper-Kamiokande) detector.}

\FullConference{%
  40th International Conference on High Energy physics - ICHEP2020\\
  July 28 - August 6, 2020\\
  Prague, Czech Republic (virtual meeting)
}

\begin{document}
\maketitle

\section{Introduction}

The discovery of neutrino oscillations, a consequence of neutrino masses and mixing between the three standard model generations of leptons, is milestone of particle physics. Three flavors of ``active'' neutrinos are consistent with mixing angle results within error bars, however, there are anomalies may be hints of physics beyond the standard model (see, e.g., refs. \cite{Kopp:2013vaa,Giunti:2013aea} and references therein), although there are tensions between results from appearance and disappearance experiments. The anomalies  can accommodate a fourth mostly ``sterile" neutrino species with an eV-scale mass and a mixing angle with active neutrinos of order $\theta_0\simeq 0.1$ \cite{Kopp:2013vaa,Giunti:2013aea}.
Sterile neutrinos by definition have no direct coupling to the weak gauge bosons, however, they may interact via a ``secret interaction" mediator in the sterile sector.  It is this possibility that is described here, in the context of the diffuse supernova neutrino background (DSNB) flux
\cite{Jeong:2018yts}. In the discussion below, for simplicity, we denote the mass eigenstates as active and sterile neutrinos, although these states have components of all four flavor states.

Supernovae are sources of neutrinos. The diffuse supernova neutrino background flux peaks at neutrino energies of a few MeV, with a flux extending to 10's of MeV. In a $3+1$ active-sterile neutrino scenario, cosmic eV-scale relic sterile neutrinos together with 10 MeV supernovae neutrinos can probe keV-scale gauge boson mediators of secret interactions. Spectral dips in the diffuse SN neutrino flux, analogous to the spectral dips in an ultrahigh energy neutrino flux traveling through the cosmic SM neutrino background, will be pronounced. DUNE and Hyper-Kamiokande measurements of  these spectral dips would probe physics from beyond the standard model. 

We review cosmological constraints on sterile neutrinos with secret interactions to show that for sterile neutrino mass, mixing and secret interactions characterized by an interaction Lagrangian ~\cite{Dasgupta:2013zpn, Chu:2015ipa}
\begin{equation}
{\cal L}_s = g_s\bar{\nu}_s\gamma_\mu P_L\nu_s\phi^\mu\ ,
\end{equation}
for vector boson mediator $\phi$ with mass $M_\phi$ and coupling $g_s$, with $P_L=(1-\gamma_5)/2$, the range of $M_\phi=4-8$ keV and $g_s\simeq 10^{-4}-5\times 10^{-3}$ is permitted for sterile neutrino mass $m_s=1$ eV and $\theta_0=0.1$. A key feature with keV-scale gauge bosons is that for most  processes, a four-fermion effective interaction is not a good approximation to the sterile neutrino interactions.  We show the impact of absorption on the diffuse supernova neutrino spectra
from sterile neutrino interactions on cosmic relic sterile neutrinos. We discuss the potential for DUNE and Hyper-Kamiokande to observe these spectral distortions.

\section{Cosmological constraints}

Cosmological constraints discussed here come from the epoch of Big Bang nucleosynthesis (BBN) and from free-streaming constraints.

The nucleosynthesis of the light elements depends on the expansion of the Universe when the photon temperature is $T_\gamma\sim 1$ MeV, the temperature at the time of nucleosynthesis, so the number of relativistic degrees of freedom at that temperature can be constrained. The limit on the number of relativistic degrees of freedom can be translated to a limit on the effective number of neutrinos $N_{\rm eff}^{\rm BBN}$. The BBN limit is \cite{Cyburt:2015mya}
\begin{equation}
N_{\rm eff}^{\rm BBN}\lsim 3.2\ .
\end{equation}
The application of this constraint depends on the ratio of the sterile and active neutrino temperatures. We assume that the sterile neutrinos and $\phi$ decouple at the TeV scale. In the absence of oscillations, when $M_\phi\gsim 1$ MeV, the $\phi$ is non-relativistic during BBN. The effective number of neutrino species that accounts for both the $\phi$ and sterile neutrino is $N_{\rm eff}^{\rm nr}\simeq 3.22$. When the $\phi$ is relativistic, the effective number of degrees of freedom is 
$N_{\rm eff}^{\rm rel}\simeq 3.17$. Our conclusion is that in the absence of neutrino flavor mixing between active and sterile neutrinos, the BBN constraint is satisfied. 

Since we assume the sterile neutrinos do mix with active neutrinos, we require that active-sterile neutrino conversions occur at a rate $\Gamma_{\nu_s}(\nu_a\to\nu_s)$ that is less than the Hubble expansion rate $H$ for $T_\gamma>1$ MeV. The active sterile conversion rate depends on the average probability for active-sterile conversions $\langle P(\nu_a\to \nu_s)\rangle$ and the quantum damping rate $D_{\rm int}= \Gamma_{\rm int}/2=(\Gamma_{\rm int,SM}+
\Gamma_{\rm int,s})/2$ that accounts for collisions:
\begin{eqnarray}
\nonumber 
\Gamma_{\nu_s}(\nu_a\to\nu_s) &=& \frac{\Gamma_{\rm int}}{2}\times \langle P(\nu_a\to \nu_s)\rangle
\\
&\simeq& D_{\rm int} \times \frac{1}{2}\frac{(\Delta m^2_s/2E)^2\sin^2 2\theta_0}{((\Delta m^2_s/2E)\cos 2\theta_0 +V_{\rm eff})^2+
(\Delta m^2_s/2E)^2\sin^2 2\theta_0+D_{\rm int}^2}\ ,
\end{eqnarray}
where $\Delta m_s^2=1$ eV$^2$ is assumed here. In medium mixing 
given the effective potential $V_{\rm eff}$ reduces the vacuum mixing angle $\theta_0$ significantly so that $\langle P(\nu_a\to \nu_s)\rangle$ is sufficiently small for a range of $(g_s,M_\phi$) \cite{Dasgupta:2013zpn}. The thermal average of the sterile neutrino cross section is strongly influenced over a wide range of temperatures by the resonance production of $\phi$, which feeds into the BBN constraints on the coupling constant and mediator mass.  At high temperatures, the thermal average of $t$-channel $\nu_s\nu_s$ interaction cross section times relative velocity is constant, as emphasized in refs. \cite{Cherry:2014xra,Cherry:2016jol}. 

At much lower temperatures, for $T_\gamma < 1$ eV, we also require that $\Gamma_{\nu_s}/H<1$. In the standard model, active neutrinos are free streaming when $T_\gamma = 1$ eV, and their free streaming affects evolution of matter density and cosmic microwave background fluctuations.  

\begin{figure}[htbp]
\begin{center}
\includegraphics[width=0.49\columnwidth]{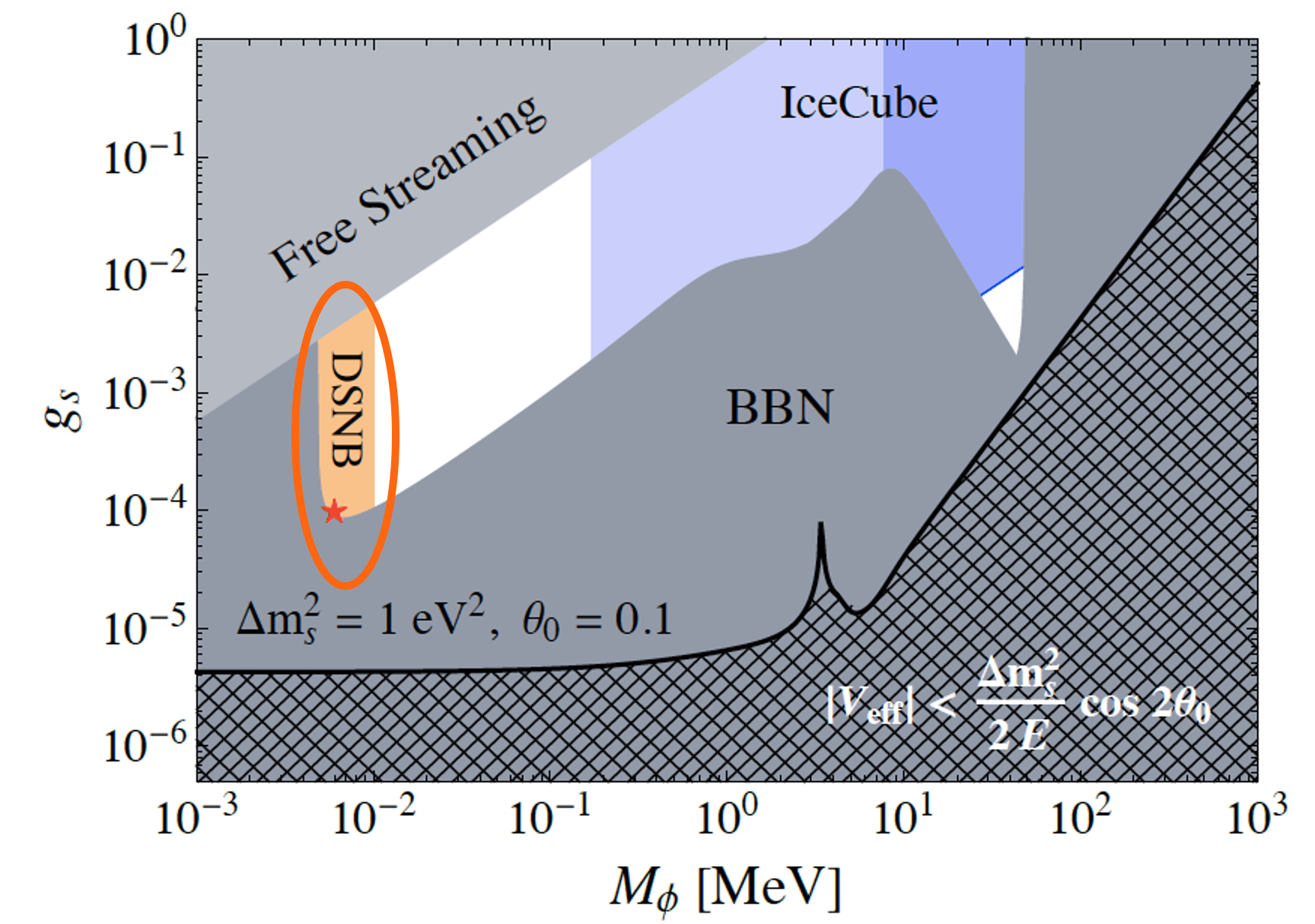}
\caption{Allowed and excluded regions of the secret interaction coupling $g_s$ and vector boson mediator mass $M_\phi$. The gray regions are excluded by free-streaming and Big Bang Nucleosynthesis (BBN) constraints. The cross-hatched region gray region shows where active-sterile neutrino mixing can be determined neglecting $V_{\rm eff}$.
The orange allowed region is relevant for this work, where the star indicates our default parameter choice of $g_s=10^{-4}$
and $M_\phi=6$ keV. The blue region shows where  IceCube can in principle constrain
secret interactions with MeV scale $M_\phi$ using astrophysical neutrinos  scattering with relic active neutrinos (light blue) and sterile neutrinos (darker blue). }
\label{fig:cosmolimits}
\end{center}
\end{figure}

In Fig. \ref{fig:cosmolimits}, the gray shaded regions are excluded. The shaded region labeled by ``Free Streaming'' shows the parameter space where the required decoupling of sterile neutrinos by the epoch when $T_\gamma=1$ eV is not met, thus do not satisfy CMB epoch cosmological constraints. BBN constrains smaller $g_s$ as shown in the figure. 
The colored and white regions of  Fig. \ref{fig:cosmolimits} show the parameter space that is allowed for $\Delta m_s=1$ eV$^2$. The circled orange region of parameter space has values of $(g_s,M_\phi)$ that are, in principle, accessible to DUNE and Hyper-Kamiokande through the DSNB. The starred point is our default parameter choice, $g_s=10^{-4}$ and $M_\phi=6$ keV. IceCube can test MeV scale $M_\phi$ with astrophysical neutrinos in the parameter space shaded in blue.

\section{Diffuse supernova neutrino background flux with secret interactions}

Secret interactions influence the DSNB flux that arrives at Earth. Since active neutrinos are detected, we use the active 
neutrino survival probability $P_i(E_\nu , z) = e^{- \tau_i (E_\nu, z)}$, written in terms of the optical depth associated with resonant scattering cross section $\sigma_s$ with relic sterile neutrinos, 
\begin{equation}
\label{eq:tau}
\tau_i (E_\nu, z) \simeq \int_0^z \frac{dz'}{H(z') (1+z')} \, n_s^0 \, (1 + z')^3 \, |U_{si}|^2 \, \sigma_s (z',E_\nu) \  ,
\end{equation}
where $|U_{si}|^2$ is square of the sterile-active mixing for active eigenstate $i$. The sterile neutrino number density today 
is 
\begin{equation}
n^0_s \simeq \frac{1}{2} \, n^0_{\nu + \bar{\nu}} \, \left(\frac{T_{as}}{T_\nu}\right)^3 \simeq 51~{\rm cm^{-3}} ~,
\end{equation}
given $n^0_{\nu + \bar{\nu}} \simeq 112 \, {\rm cm^{-3}}$, and $T_{as}\simeq T_\nu$ is the sterile neutrino temperature after first recoupling with active neutrinos, then subsequent sterile neutrino decoupling while they are still relativistic.

The redshift-integrated DSNB flux not absorbed by the 
resonance interaction is
\begin{equation}
\label{eq:nufluxEarth-dsnb-res}
F_a (E_\nu) =  \sum_{i = 1}^4 \left| U_{ai}\right|^2 \, \int_{0}^{z_{\rm max}} dz \, P_i(E_\nu , z) \, R_{\rm SN} (z) \, F_i^0(E') \, (1+z) \, \left|\frac{dt}{dz} \right|  ~,
\end{equation}
where $E'=(1+z)E_\nu$ is the neutrino energy emitted at redshift $z$, $ F_i^0(E')$ is the
spectrum of neutrinos of mass eigenstate $i$ exiting the supernova  after adiabatic propagation in the interior \cite{Keil:2002in,Lunardini:2010ab,Esmaili:2014gya}, $R_{\rm SN}(z)$ is the cosmic SN rate
\cite{Horiuchi:2011zz} and
\begin{equation}
\label{eq:dzdt}
\frac{dt}{dz} = - \left( H_0 (1+z) \sqrt{\Omega_m (1+z)^3 + \Omega_\Lambda}\right)^{-1} ~,
\end{equation}
for $\Omega_m = 0.308 \pm 0.012$,  $\Omega_\Lambda = 0.692 \pm 0.012$ and $H_0 = (67.8 \pm 0.9) \, {\rm km \ s^{-1}  Mpc^{-1}}$~\cite{Ade:2015xua}. 
The sterile-active sector mixings in the four-flavor mixing matrix $U_{ai}$ are assumed to be 
$\theta_{14}=\theta_{24}=\theta_{34}\equiv\theta_0 = 0.1$.  In Ref. \cite{Jeong:2018yts}, we evaluate results for the normal hierarchy (NH) and the inverted mass hierarchy, with high energy (HE) and low energy parameter sets in the description of the neutrino spectrum of core collapse supernovae \cite{Lunardini:2010ab}. We illustrate our results here with the NH in the more optimistic HE parametrization of the neutrino spectrum.

\begin{figure}[htbp]
\begin{center}
	\includegraphics[width=0.49\textwidth]{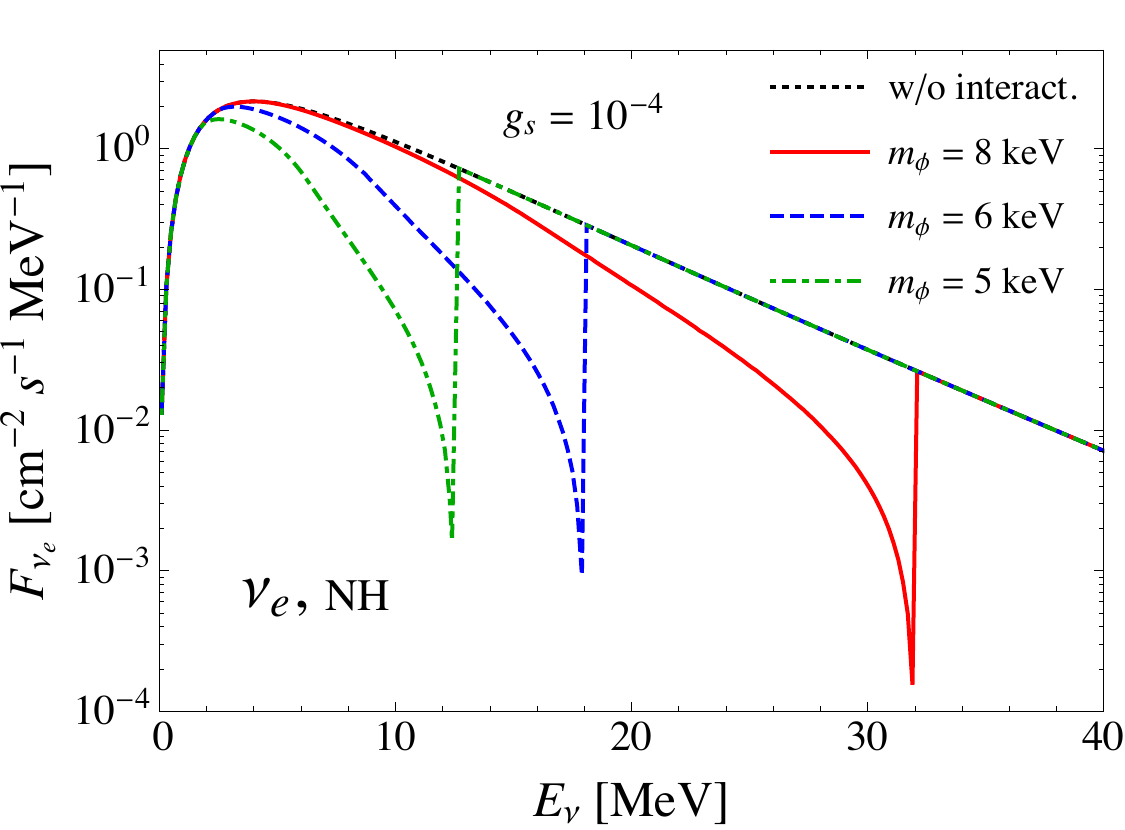}
	\includegraphics[width=0.49\textwidth]{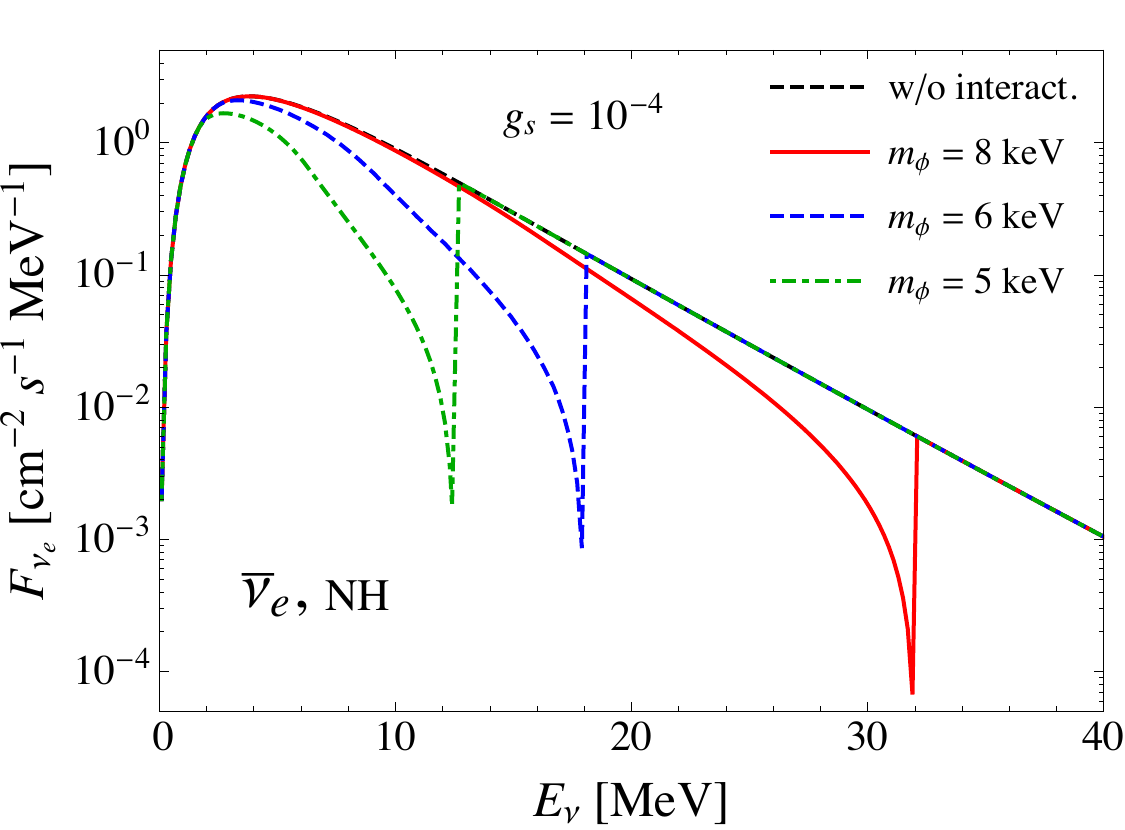}
\caption{The $\nu_e$ and $\bar{\nu}_e$ DSNB flux evaluated with the HE flux parameters of Ref. \cite{Lunardini:2010ab} for neutrinos in the normal mass hierarchy, without interactions (black dashed) and with interactions with relic sterile neutrinos and mediator masses $M_\phi=5,\ 6$ and $8$ keV and coupling $g_s=10^{-4}$, as in Ref. \cite{Jeong:2018yts}.}
\label{fig:flux}
\end{center}
\end{figure}

\begin{figure}[htbp]
\begin{center}
	\includegraphics[width=0.49\textwidth]{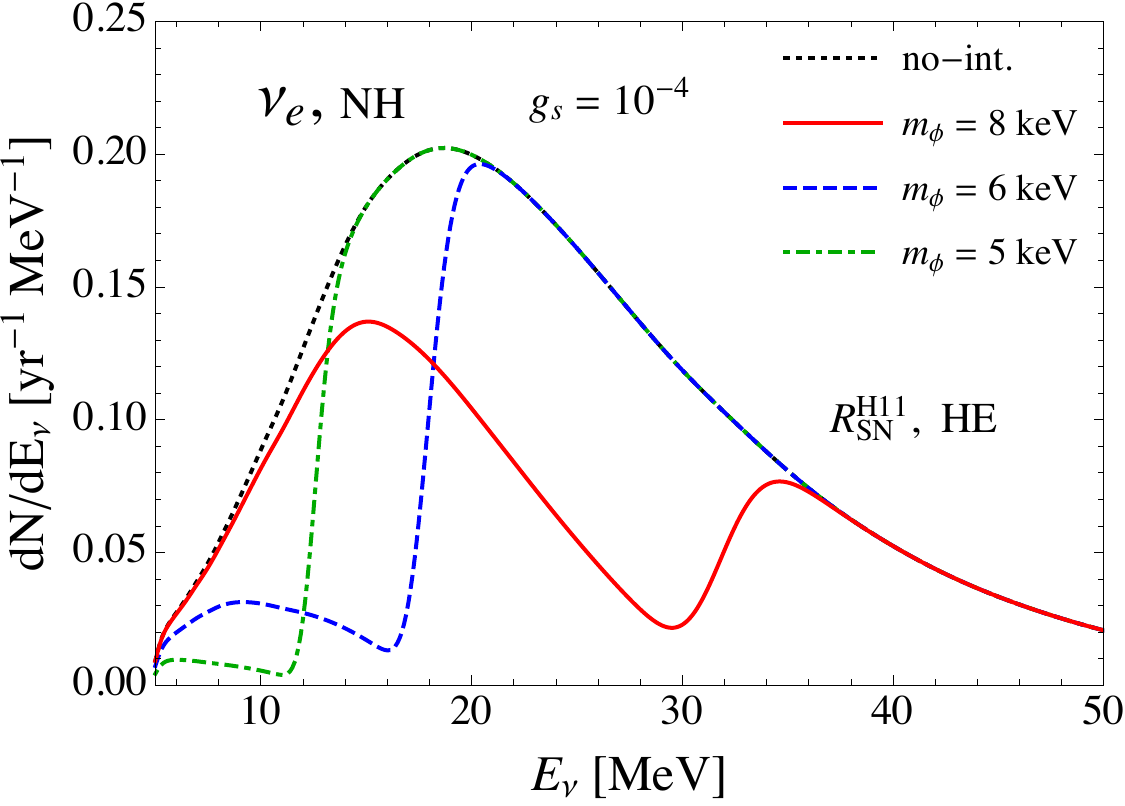}
	\includegraphics[width=0.49\textwidth]{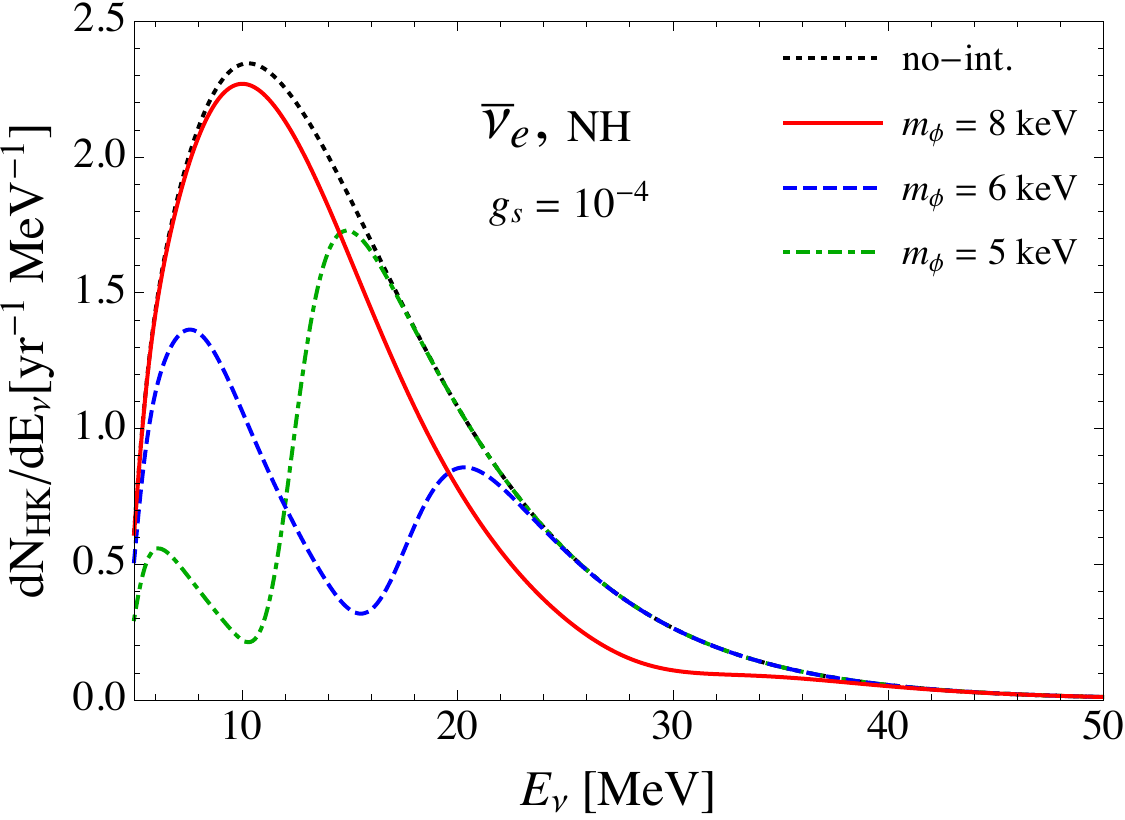}
\caption{For the same parameters in Fig. 2, the $\nu_e$ differential event rates for 40 kton DUNE LAr detection of $\nu_e$ CC events (left) and
for $\bar{\nu}_e$ inverse beta decay events in 187 kton of water at Hyper-Kamiokande, as in Ref. \cite{Jeong:2018yts}. }
\label{fig:events}
\end{center}
\end{figure}

Figure \ref{fig:flux} shows the $\nu_e$ (left) and $\bar{\nu}_e$ (right) DSNB fluxes with and without resonance interactions in transit through the relic sterile neutrino background. The neutrino fluxes are convoluted with detector energy resolution functions and reaction cross sections for DUNE LAr and Hyper-Kamiokande water detection of $\nu_e$ and $\bar{\nu}_e$ interactions, respectively, described in detail in Ref. \cite{Jeong:2018yts}. 
As Fig. \ref{fig:events} shows, the sharp absorption dips are smeared by detection effects, and energy dependence of the differential event rates reflects the respective (anti-)neutrino interaction cross sections.

\section{Conclusions}

The double-peaked structure of the differential event rates comes from the rising cross section and the falling neutrino spectrum as the neutrino energy increases, together with the absorption dip. Solar neutrinos are a significant background for $E_\nu\lsim 16$ MeV. If $M_\phi\sim 8-9$ keV, the double-peak structure may be detectable if the energy threshold for detection is low enough, given enough events and an understanding of the solar neutrino spectrum in this energy range. For $16$ MeV$\leq E_\nu\leq 40$ MeV, the NH with $M_\phi=8$ keV yields 16 events after 10 years of operation with a 400 kton$\cdot$yr LAr detector. This compares to 32 events without $\nu_s$ and 32 events with 3 active and 1 sterile flavor, but where secret interactions do not occur with a resonance in the $\sim 5-10$ keV mass region.
Uncertainties in the inputs to the calculation include uncertainties in the neutrino cross section, the supernova energy spectra, and the supernova formation rate. It is possible that the number of events could be an order of magnitude larger. 
The number of $\bar{\nu}_e$ events for 2.6 Mton$\cdot$yr for 10 years of Hyper-Kamiokande yields a wider range of predictions: 121 events in the NH with $M_\phi=8$ keV, 179 events with 3 active and 1 sterile flavor but no relevant resonant interactions, and 316 events in the three active flavor scenario. Similar results are obtained with the inverted hierarchy.

This work was supported in part by the US Department of Energy (DE-SC-0010113, DE- SC-0010114, DE-SC-0002145 and DE-SC0009913), the Spanish MINECO (FPA2017-84543-P), the EU Horizon 2020 (H2020-MSCA-ITN-2019//860881-HIDDeN) and the Portuguese FCT (UID/FIS/00777/2019 and CERN/FIS-PAR/0004/2019).

\end{document}